\documentclass[twocolumn]{revtex4}
\usepackage[dvips]{graphicx}
\usepackage{dcolumn}
\usepackage{bm}

\begin{document}

\title{Transport in graphene nanostructures with spatially modulated gap and potential}

\author{E.S.~Azarova}
\author{G.M.~Maksimova}
\altaffiliation[Corresponding author. ]{Tel.: +7 831 4623304; \\
E-mail address: maksimova.galina@mail.ru (G.M. Maksimova)}

\address{Department of Theoretical Physics, University of Nizhny Novgorod, 23 Gagarin Avenue, 603950 Nizhny Novgorod, Russian Federation}

\date{\today}

\begin{abstract}

We study transport properties of graphene nanostructures consisted of alternating slabs of gapless $(\Delta=0)$ and gapped $(\Delta\ne 0)$ graphene in the presence of piecewise constant external potential equal to zero in the gapless regions. The transmission through single-, double-barrier structures and superlattices has been studied. It was revealed that any $n$-barrier structure is perfectly transparent at certain conditions defining the positions of new Dirac points created in the superlattice. The conductance and the shot noise were as well computed and investigated for the considered graphene systems. In a general case, existence of gapped graphene fraction leads to decrease of the conductance and increase of the Fano factor. For two barriers formed by gapped graphene and separated by a long and highly doped region the Fano factor rises up to~0.5 in contrast to the similar gapless structure where the Fano factor is close to~0.25. Similarly to a gapless graphene superlattice, creation of each new Dirac point manifests itself as a conductivity resonance and a narrow dip in the Fano factor. However, gapped graphene inclusion into the potential-barrier regions in the superlattice leads to more complicated dependence of the Fano factor on the potential height compared to pseudo-diffusive behaviour (with $F=1/3$) typical for a gapless superlattice.

\end{abstract}

\maketitle

\section{Introduction}

Transport properties of graphene and graphene-based microstructures are currently among the most actively investigated topics in graphene physics~\cite{Sarma,Castro,Peres,Mucc,Kats,Tworz,DiCarlo,Dann,Balandin,Ghosh,Novoselov}
. Aside from fundamental aspects such interest in graphene stems from its potential applications as a high-mobility semiconductor and the experimental ability to tune its properties via gating~\cite{Castro}
. Investigation of the electron transport includes the consideration of a conductance and shot noise which is characterized by the Fano factor $F$ being the ratio of the noise power and mean current. For instance, the Fano factor of wide and short graphene sheet equals 1/3~\cite{Tworz} 
near the Dirac point. This coincides with the well-known result for diffusive wire~\cite{Beenakker}
.

Lots of theoretical and experimental works have been devoted to investigations of transmission $T$ and conductance $G$ through different multibarrier graphene nanostructures and graphene superlattices (SLs)~\cite{Wang,Huard,Bliokh,Krst,Pell,Gatt,Pereira,Brey,Park,Meyer,Mar} 
which can be fabricated, e.g., by applying a local top gate voltage. It has been shown that a one-dimensional periodic potential substantially affects the transport properties of graphene. For instance, the Kronig-Penny type electrostatic potential produces strong anisotropy in the carrier group velocity near the Dirac point leading to the supercollimation phenomenon~\cite{Park1,Park2,Barbier}
.

The band structure of an ideal graphene sheet has no energy gap which results, for example, in total transparency of any potential barrier for normally-incident electrons~\cite{Kats1} 
(an analog of the Klein paradox~\cite{Klein}
). It is extremely desirable for electronics applications that graphene structures be gapped. Therefore, much effort of researchers has been focused on producing a gap in the graphene spectrum. The gap can be created by strain engineering as well as by deposition or adsorption of molecules on a graphene layer. For instance, a hydrogenated sheet of graphene (graphane) is a semiconductor with a gap of the order of a few~eV~\cite{Leb}
. Other way of producing the gap is to use hexagonal boron nitride (hBN) substrate. In this case the gap value is small enough owing to the lattice mismatch. However, it can be increased by the applying of a perpendicular electric field~\cite{Kind}.

Creation of various graphene heterostructures, including SLs, with the gap discontinuity is widely discussed now. One way of generating spatially modulated gap is graphene on a substrate made from different dielectrics~\cite{Ratn}
. The required gap modulation can also be created by using, e.g., an inhomogeneously hydrogenated graphene or graphene sheet with nonuniformly deposited CrO$_3$ molecules. In our previous work~\cite{Maks} 
we studied the electronic properties of graphene SL in which the gap and potential profile are piecewise constant functions. It was found that in such SL up to some critical value $V_c$ of potential allowed subbands are separated by gaps. When the potential value is greater or equal to $V_c$ the contact or cone-like Dirac points appear in the spectrum. As a result, SL becomes gapless.

In this work we examine in detail ballistic transport through graphene nanostructures, including SL, formed by space-modulated gap and potential. Using the transfer-matrix formalism we study the transmission, conductance and the Fano factor for systems with arbitrary numbers of barriers.

\section{Basic equations}

Let us initially consider a lateral one-dimensional multibarrier structure consisting of $N$ strips with widths $d_j$ $(j=1,\dots,N)$characterized by the gaps $\Delta_j$ and potential heights $V_j$ (see Fig.~\ref{fig1}). The outer regions labeled by 0 and $N+1$ correspond to the gapless graphene with $\Delta=V=0$. In $j^{th}$ strip, the carriers are described by the two-dimensional Dirac equation
\begin{eqnarray}\label{eq1}
(\hbar\upsilon_F\boldsymbol {\sigma}{\bf k}+\Delta_j\sigma_z)\Psi_j=(E-V_j)\Psi_j,
\end{eqnarray}
where $\hbar{\bf k}$ is the momentum operator, $\boldsymbol {\sigma}$ is the vector of Pauli matrices, and $\upsilon_F\approx 10^6m/s$ is the Fermi velocity. Due to translation invariance in the $y$-direction, the solution of Eq.~(\ref{eq1}) in $j^{th}$ region can be written as $\Psi_j(x,y)=\Psi_j(x)\exp (ik_yy)$. It is convenient to define the wavevector $k_j$ as
\begin{eqnarray}\label{eq2}
k_j=\frac{\sqrt{(E-V_j)^2-\Delta^2_j}}{\hbar\upsilon_{F}}.
\end{eqnarray}
Then for $k_j^2>k_y^2$ the wavefunction $\Psi_j(x)$ in strip $j$ $(x_j^L\le x\le x_j^R)$ is a superposition of plane waves
\begin{eqnarray}\label{eq3}
&&\Psi_j(x)=\frac{A_j}{\sqrt{\delta_j^2+1}}\pmatrix{1 \cr
\sigma_j\delta_je^{i\theta_j}}\exp(ik_jx\cos\theta_j)+ \nonumber\\
&&+\frac{B_j}{\sqrt{\delta_j^2+1}}\pmatrix{1 \cr
-\sigma_j\delta_je^{-i\theta_j}}\exp(-ik_jx\cos\theta_j).
\end{eqnarray}
Here, $\theta_j=\tan^{-1}(k_y/k_{xj})$, $k_{xj}=\sqrt{(E-V_j)^2-\Delta_j^2-(\hbar \upsilon_Fk_y)^2}/\hbar \upsilon_F$, $\theta_j\in[-\pi/2,\pi/2]$, $\delta_j=\sqrt{(E-V_j-\Delta_j)/(E-V_j+\Delta_j)}$, $\sigma_j=\text{sgn}(E-V_j+\Delta_j)$. $x_j^L$ and $x_j^R$ denote the left and right boundaries of the strip $j$, so that $x_{j-1}^R=x_j^L$. In the opposite case, when $k_j^2<k_y^2$, solution $\Psi_j(x)$ has pure exponential behaviour along the $x$-axis.

Suppose that $\Psi_j(x)$ oscillates everywhere. Then we define the functions $A_j(x)=A_j\exp(ik_jx\cos\theta_j)$, $B_j(x)=B_j\exp(-ik_jx\cos\theta_j)$. As a result, Eq.~(\ref{eq3}) may be written in the form
\begin{eqnarray}\label{eq4}
\Psi_j(x)=L_j\pmatrix{A_j(x) \cr B_j(x)},
\end{eqnarray}
where
\begin{eqnarray}\label{eq5}
L_j=\frac{1}{\sqrt{\delta_j^2+1}}\pmatrix{1 & 1\cr
\sigma_j\delta_je^{i\theta_j} & -\sigma_j\delta_je^{-i\theta_j}}.
\end{eqnarray}
Continuity of the upper and lower components $\Psi_j(x)$ at the strip boundaries requires that
\begin{eqnarray}\label{eq6}
L_{j-1}\pmatrix{A_{j-1}^R\cr B_{j-1}^R}=L_{j}\pmatrix{A_{j}^L\cr B_{j}^L}.
\end{eqnarray}
Within the region $j$ the solutions $(A_j^L$, $B_j^L)$ and $(A_j^R$, $B_j^R)$ are connected by the free propagation matrix $K_j$:
\begin{figure}[t] \centering
\includegraphics*[scale=0.6]{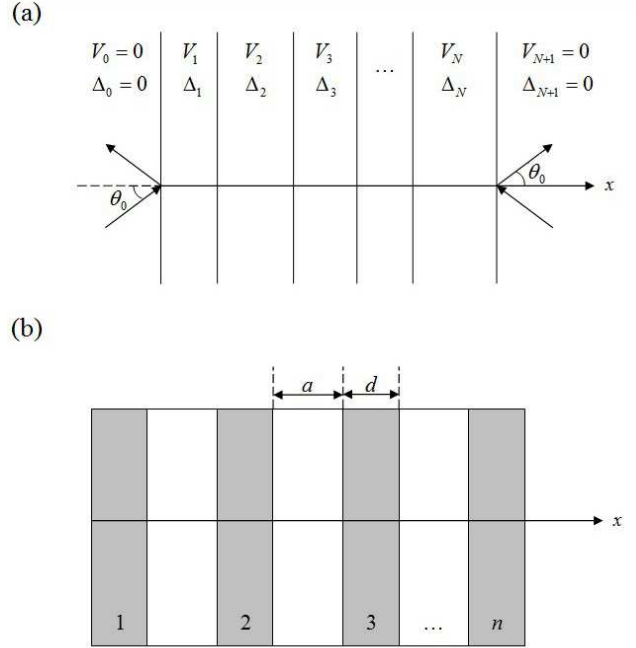}
\caption{(a) Model of graphene structure represented by series slabs of width $d_j$ ($j=1,\dots,N$) characterized by gaps $\Delta_j$ and potential $V_j$.
(b) Schematic diagram of a Kronig-Penney type multibarrier structure, in which the gap and potential equal to $\Delta$ and $V$ respectively in the grey regions and zero outside.} \label{fig1}
\end{figure}
\begin{eqnarray}\label{eq7}
\pmatrix{A_{j}^R\cr B_{j}^R}=K_{j}\pmatrix{A_{j}^L\cr B_{j}^L},
\end{eqnarray}
where
\begin{eqnarray}\label{eq8}
 K_j=\pmatrix{e^{ik_jd_j\cos\theta_j} & 0\cr 0 & e^{-ik_jd_j\cos\theta_j}}.
\end{eqnarray}
Combining Eqs~(\ref{eq7}) and (\ref{eq8}) one can find:
\begin{eqnarray}\label{eq9}
\pmatrix{A_{N+1}^L\cr B_{N+1}^L}=M\pmatrix{A_{0}^R\cr B_{0}^R},
\end{eqnarray}
where the transfer matrix $M$ is introduced for the considered heterostructure as
\begin{eqnarray}\label{eq10}
M=L_{N+1}^{-1}F_NF_{N-1}\dots F_1L_0.
\end{eqnarray}
Here $L_{N+1}=L_0$ is determined by Eq.~(\ref{eq5}) at $V=\Delta=0$ and $F_j=L_jK_jL_j^{-1}$, which yields:
\begin{eqnarray}\label{eq11}
&&F_j=\frac{1}{\cos\theta_j}\pmatrix{\cos(k_jd_j\cos\theta_j-\theta_j) & \frac{i\sigma_j}{\delta_j}\sin(k_jd_j\cos\theta_j) \cr
i\sigma_j\delta_j\sin(k_jd_j\cos\theta_j) & \cos(k_jd_j\cos\theta_j+\theta_j)}.\nonumber\\
&&
\end{eqnarray}
We may use Eq.~(\ref{eq11}) for an arbitrary multibarrier structure, characterized by different parameters $\Delta_j$ and $V_j$ in each slab of width $d_j$.

We now consider scattering of a Dirac particle on the graphene superstructure consisted of $n$ gapped graphene strips of width $d$ and $(n-1)$ gapless graphene strips of width $a$. Then $\Delta_j$ and $V_j$ equal $\Delta$ and $V$, respectively, in the gapped regions and zero elsewhere. In this case the transfer matrix $M$ can be written in the form
\begin{eqnarray}\label{eq12}
M^{(n)}=GS^{n-1}L_0,
\end{eqnarray}
where expressions for the matrices $G$ and $S$ are given in the Appendix [Eqs~(\ref{eqa4})~and~(\ref{eqa5})]. If $n$ is large enough, it is convenient to use the $S$-representation in which $S$ matrix is diagonal. The diagonalization procedure is a transform $S'=U^{-1}SU$ with
\begin{eqnarray}\label{eq13}
U=\pmatrix{1 & 1 \cr c & d},
\end{eqnarray}
where
\begin{eqnarray}\label{eq14}
c=\frac{i(s_{11}-\lambda_+)}{s_{12}}, \text{ }
d=\frac{i(s_{11}-\lambda_-)}{s_{12}}.
\end{eqnarray}
Then,
\begin{eqnarray}\label{eq15}
S'=\pmatrix{\lambda_+ & 0 \cr 0 & \lambda_-},
\end{eqnarray}
where
\begin{eqnarray}\label{eq16}
\lambda_{\pm}=\frac{s_{11}+s_{22}}{2}\pm \sqrt{\frac{(s_{11}+s_{22})^2}{4}-1}.
\end{eqnarray}
Using the above relations we obtain the final expression for the transfer matrix
\begin{eqnarray}\label{eq17}
M^{(n)}=(GU)S'^{n-1}(U^{-1}L_0).
\end{eqnarray}
Supposing that the incoming wave is scattered on the left border of the structure, we set $A_0^R=1$, $B_0^R=r$, $A_{N+1}^L=t$ and $B_{N+1}^L=0$. Here $r$ and $t$ are the amplitudes of the reflected and transmitted states. Then the transmission probability $T=|t|^2$ is given by
\begin{eqnarray}\label{eq18}
T=|M_{22}^{(n)}|^{-2}.
\end{eqnarray}
Substituting Eqs~(\ref{eq5}), (\ref{eq13})--(\ref{eq15}), and (\ref{eqa3}) into Eq.~(\ref{eq17}
), one obtains
\begin{eqnarray}\label{eq19}
&&M_{22}^{(n)}=\frac{(g_{21}+cg_{22})(d+\sigma_0e^{-i\theta_0})\lambda_+^{n-1}}{2(d-c)\cos\theta\cos\theta_0}- \nonumber\\
&&-\frac{(g_{21}+dg_{22})(c+\sigma_0e^{-i\theta_0})\lambda_-^{n-1}}{2(d-c)\cos\theta\cos\theta_0},
\end{eqnarray}
where matrix elements $g_{ij}$  are defined by Eq.~(\ref{eqa4}).

\section{Tunneling through multiple barriers}

\begin{figure}[t] \centering
\includegraphics*[scale=0.6]{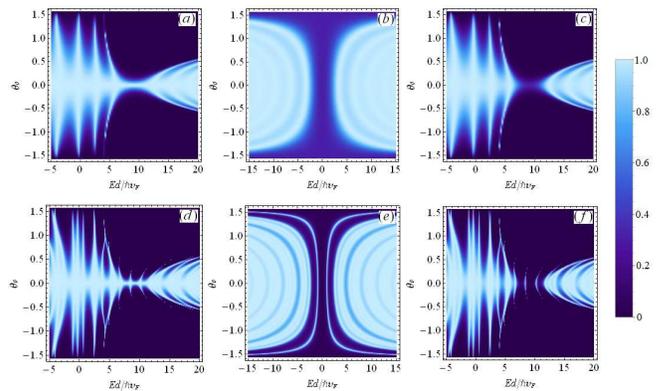}
\caption{Density plot of the transmission through a single gap-potential barrier (a,~b,~c) and for double barrier (d,~e,~f) for $Vd/\hbar\upsilon_F=9.11$, $\Delta=0$~(a,~d); $V=0$, $\Delta d/\hbar\upsilon_F=1.21$~(b,~e) and $Vd/\hbar\upsilon_F=9.11$, $\Delta d/\hbar\upsilon_F=1.21$~(c,~f).}
\label{fig2}
\end{figure}
We first consider the single-barrier geometry when the gapped graphene strip of width $d$ borders on the gapless graphene. Using Eq.~(\ref{eqa6}) one has
\begin{eqnarray}\label{eq20}
&&T(E,\theta_0)=\left[1+\left(\frac{\Delta^2\cos^2\theta_0+V^2\sin^2\theta_0}{(\hbar \upsilon_Fk_x)^2\cos^2\theta_0}\right)\sin^2k_xd\right]^{-1}.\nonumber\\
&&
\end{eqnarray}
Here the wave vector $k_x$ is defined in the barrier region and also depends on the angle of incidence $\theta_0$ (or $k_y=|E|\sin\theta_0/\hbar\upsilon_F$):
\begin{eqnarray}\label{eq21}
k_x=\sqrt{(E-V)^2-\Delta^2-E^2\sin^2\theta_0}/\hbar \upsilon_F.
\end{eqnarray}
Eq.~(\ref{eq20}) is a generalization of two cases: $\Delta=0$, $V\ne 0$~\cite{Kats2} 
and $\Delta\ne 0$, $V=0$~\cite{Gomes}. 
The transmission described by Eq.~(\ref{eq21}) oscillates as a function of barrier width $d$ with a period depending on the wavevector $k_y$. Such behaviour takes place at all initial angles $\theta_0$ when the particle energy $E<E_0$, where
\begin{eqnarray}\label{eq22}
E_0=\frac{V^2-\Delta^2}{2V}.
\end{eqnarray}
Note that, at $E=E_0$ the wave vector $k_j$ is the same for both gapped and gapless regions. If the energy satisfies the conditions $E_0<E<V-\Delta$ or $E>V+\Delta$ similar oscillating dependence holds only for the angles of incidence $\theta_0<\theta_{0c}$, where
\begin{eqnarray}\label{eq23}
\theta_{0c}=\sin^{-1}\sqrt{((E-V)^2-\Delta^2)/E^2}.
\end{eqnarray}
When $\theta_0$ exceeds the critical angle $\theta_{0c}$, $k_x$ is pure imaginary and the transmission is determined by the evanescent states in barrier region
\begin{eqnarray}\label{eq24}
T(E,\theta_0)=\left[1+\left(\frac{\Delta^2+V^2\tan^2\theta_0}{(\hbar \upsilon_F\kappa)^2}\right)\sinh^2\kappa d\right]^{-1},
\end{eqnarray}
where
\begin{eqnarray}\label{eq25}
\kappa=\sqrt{E^2\sin^2\theta_0+\Delta^2-(E-V)^2}/\hbar v_F.
\end{eqnarray}
These expressions also describe the transmission through the barrier at all angles $\theta_0$ for the energies lying inside the gap: $V-\Delta<E<V+\Delta$. To illustrate the dependence of the transmission on both the energy and the angle of incidence we construct a density plot of $T$. The different colors from black to white correspond to different values of $T$ from 0 to 1. Such a density plot for a single barrier of width $d = 30$ nm is shown in figures~\ref{fig2}(a)--\ref{fig2}(c) at various ratios between $V$ and $\Delta$. For a gapless graphene it is clearly seen (Fig.~\ref{fig2}(a)) the perfect transmission $(T = 1)$ for normal or near-normal incidence $(\theta_0\to 0)$, which is a manifestation of Klein tunneling. The opening gap in the barrier region suppresses this effect (figures~\ref{fig2}(b),~\ref{fig2}(c)). The barrier becomes also completely transparent for values $k_xd=\pi m$, where $m$ is integer. As follows from Eq.~(\ref{eq21}) these resonances are well-defined at $\theta_0$ close to $\pi/2$. Corresponding resonant energies weakly depend on the gap value for $\Delta/V\ll1$ (figures~\ref{fig2}(a),~\ref{fig2}(c)). On the contrary, when applied potential $V=0$ and $\Delta$ is not too large $(\Delta d/\hbar\upsilon_F\approx1)$, the transmission probability is about~1 in the wide region of $E$; $\theta_0$ plane (Fig.~\ref{fig2}(b)).
\begin{figure}[t] \centering
\includegraphics*[scale=0.85]{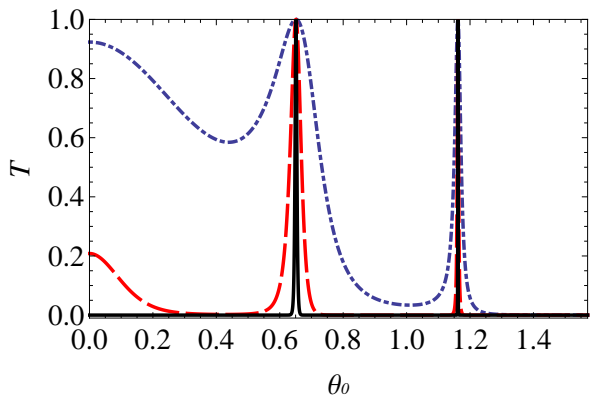}
\caption{Transmission probability as a function of angle of incidence for symmetrical multibarrier structures with $a=d=30$~nm for $V=354$~meV, $\Delta=50$~meV and $E=E_0=173$~meV: $n=1$ (dash-dotted line), $n=5$ (dashed line), $n=30$ (solid line).}
\label{fig3}
\end{figure}

To calculate the transmission through a double-barrier structure $(n=2)$ we use the expression for the real and imaginary parts of $M_{22}$ (\ref{eqa7}), (\ref{eqa8}). The results are illustrated in figures~\ref{fig2}(d)--\ref{fig2}(f) for a symmetrical case when the barrier width $d$ coincides with the interbarrier separation $a$. Pronounced resonant structure is seen in the energy interval $V-\Delta<E<V+\Delta$ caused by the quasibound states in the well region.

Note, that for $n$ identical barriers the matrix element $M_{22}^{(n)}$ (\ref{eq20}) depends on $n$ through the factors $\lambda_{\pm}$, where the eigenvalues $\lambda_+$ and $\lambda_-$ of the matrix $S$ define the band structure of infinite periodic SL with period $l$ $(l=a+d)$~\cite{Maks} 
according to equation $2\cos Kl=Sp(S)$, where $K$ is the Bloch wavevector. Thus, the infinite periodic structure is transparent when $|\lambda_{\pm}|=1$. As follows from Eq.~(\ref{eq19}), arbitrary $n$-barrier structure becomes perfectly transparent for some angles of incidence $\theta_{0m}$ in special case when $E=E_0$ and $d|E_0|/\hbar\upsilon_F\ge\pi$. Here the resonant angles $\theta_{0m}$ are given by
\begin{figure}[t] \centering
\includegraphics*[scale=0.85]{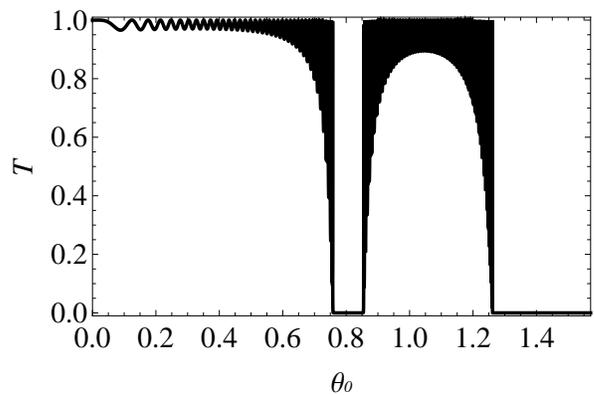}
\caption{Transmission probability as a function of angle of incidence for graphene superlattice ($a=d=30$~nm, n=30) for $V=8$~meV, $\Delta=50$~meV and $E=E_0=-149$~meV.}
\label{fig4}
\end{figure}
\begin{eqnarray}\label{eq26}
\cos\theta_{0m}=\pi\hbar \upsilon_Fm/d|E_0|, \text{ }m=1,2 \dots.
\end{eqnarray}
We should emphasize also that, the above conditions (\ref{eq22}) and (\ref{eq26}) for existence of resonances in the transmission probability of $n$-barrier structure correspond to the positions of new cone-like Dirac points in ${\bf k}$-space in the infinite SL~\cite{Maks}.

Angular dependence of the transmission coefficients $T^{(n)}(E_0,\theta_0)$ is shown in Fig.~\ref{fig3} for $n=1, \text{ }5,\text{ } 30$ and $V = 354$~meV, $\Delta=50$~meV. As seen, the positions and number of resonant peaks are defined by $E_0$ and do not change with increasing the number of barriers $n$. On the contrary, the widths of resonances decrease as $n$ increases. As a result, graphene superlattice becomes opaque for almost all angles of incidence $\theta_0$ except for $\theta_0\sim\theta_{0m}$ (see Fig.~\ref{fig3} for $n = 30$). Such a dependence $T^{(n)}(E_0,\theta_0)$ with $n\gg1$ is similar to the dependence of the transmission of electromagnetic waves in the periodic structure made of dielectric layers with refractive indices $n_1=-n_2$~\cite{Bliokh,Wu}
. The signs $\pm$ correspond, respectively, to dielectrics with positive (right-handed ($R$)) and negative (left-handed ($L$)) refractive indices.

The analogies between the charge transport in graphene structures and propagation of light in layered dielectric media have been discussed earlier~\cite{Bliokh,Che,Bliokh1}
. It was shown that the difference $E-V$ in a gapless graphene plays the same role as the refractive index in dielectric structure. In particular, focusing the electric current by a single $p-n$ junction in graphene was predicted, which is similar to focusing the electromagnetic waves by the $R$-$L$ interface~\cite{Che}
. In our case (Fig.~\ref{fig3}) the states with $E=E_0>0$ belong to the conduction band in gapless region and to the valence band $(E_0<V-\Delta)$ in gapped region, so that the considered superlattice is similar to the symmetric $R$-$L$ periodic dielectric structure. Note also that the existence of gapped fraction in graphene leads to suppression of Klein tunneling. This means that analogous $R$-$L$ periodic structure $(n_1=-n_2)$ is characterized by different impedances. When $E_0$ is negative, the angular dependence of $T^{(n)}(E_0,\theta_0)$ drastically changes (Fig.~\ref{fig4}). As seen, there are many angles other than $\theta_{0m}$, for which the transmission is also one. In this case the graphene multibarrier structure has transport properties resembling to the transmission of light through a stack of dielectric layers with the same refractive indices and different impedances~\cite{Bliokh}.

\section{Conductance and shot noise}

\begin{figure}[t] \centering
\includegraphics*[scale=0.8]{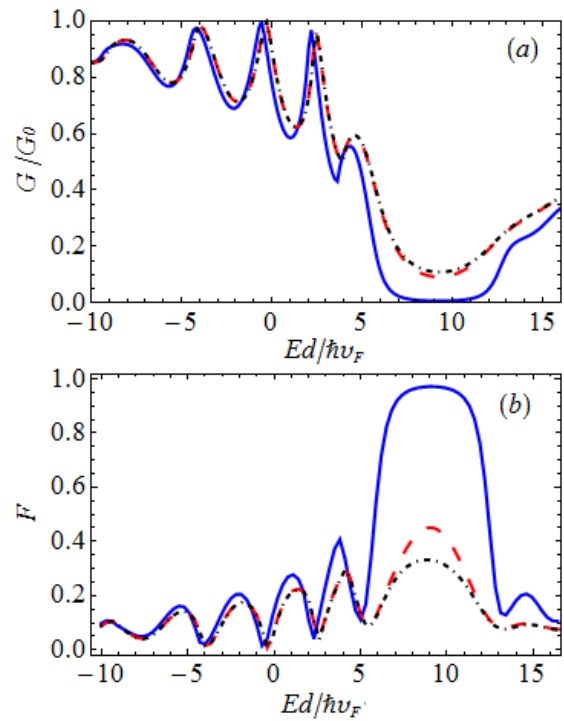}
\caption{Conductance (a)  and Fano factor (b) versus the Fermi energy for a single potential-gap barrier of width $d=30$~nm for $V=200$~meV and $\Delta=50$~meV (solid line), $\Delta=10$~meV (dashed line) and $\Delta=0$ (dash-dotted line).
} \label{fig5}
\end{figure}

Basing on the obtained results for the transmission probabilities $T(E,\theta_0)$, one can find the two-terminal Landauer conductance $G$ and the Fano factor $F$ for the finite periodic-potential-gap structure. Within a linear regime on bias voltage at very low temperatures they are given by
\begin{eqnarray}\label{eq27}
G(E)=G_0(E)\int_0^{\pi/2}{T(E,\theta_0)\cos\theta_0d\theta_0},
\end{eqnarray}
\begin{eqnarray}\label{eq28}
F(E)=\frac{\int_0^{\pi/2}{T(E,\theta_0)(1-T(E,\theta_0))\cos\theta_0 d\theta_0}}{\int_0^{\pi/2}{T(E,\theta_0)\cos\theta_0 d\theta_0}}
\end{eqnarray}
with $G_0=2ge^2EL_y/h^2\upsilon_F$ and $L_y$ the length of the slab in the $y$-direction. $g$ equals~4 due to the twofold spin and valley degeneracy. In Fig.~\ref{fig5} we plot the conductance~(a) and the Fano factor~(b) versus the Fermi energy for a single potential-gap barrier of width $d=30$~nm for $V=200$~meV and $\Delta=50$~meV (solid line), $\Delta=10$~meV (dashed line), and $\Delta=0$ (dash-dotted line). In the considered case for $Vd/\hbar\upsilon_F=9.11$ we model the leads (the gapless region) by highly doped graphene.

At $\Delta=0$ the conductance minimum and the Fano factor at the Dirac point (at $E=V$ ) coincide with the ones obtained by Tworzydlo {\it et al}.~\cite{Tworz}
\begin{eqnarray}\label{eq29}
G=G_0\hbar\upsilon_F/Vd,\mbox{ }\mbox{ }\mbox{ } F=1/3.
\end{eqnarray}
With the gap increasing, the minimum value of the conductance decreases while the maximum value of the Fano factor approaches~1. Inside the gap i.e. for $8.66<Ed/\hbar\upsilon_F<9.57$ (for $\Delta=10$~meV) and $6.83<Ed/\hbar\upsilon_F<11.39$ (for $\Delta=50$~meV) the dependencies $G(E)$ and $F(E)$ become almost flat (Fig.~\ref{fig5}). When $Vd/\hbar\upsilon_F\gg1$ and $\Delta d/\hbar\upsilon_F>1$ one can find the approximate expressions for $G(E=V)$ and $F(E=V)$:
\begin{eqnarray}\label{eq30}
G=4G_0(\hbar \upsilon_F/Vd)mK_1(2m),
\end{eqnarray}
\begin{eqnarray}\label{eq31}
F=\tanh^2m,
\end{eqnarray}
where $K_1(x)$ is the modified Bessel function of argument $x$ and $m=\Delta d/\hbar\upsilon_F$.

The results discussed above were obtained at small bias voltage between the leads and the sheet. In this case the main contribution to the current and shot noise comes from the evanescent states. At high voltages we have take into account the propagating waves also. This leads to increase of the conductance and decrease of the Fano factor. Specifically, for a gapless graphene sheet the Fano factor drops from~1/3 at low voltages to~0.125 at high voltages~\cite{Sonin}
. Thus, we may suppose, that for $\Delta\ne0$, the conductance and the Fano factor are nearly independent of finite value of voltage drops $Vd$ up to $eVd\approx\Delta$.

We have studied the double-barrier structure formed by two gapped graphene regions of width $d$ separated by highly doped region $(Vd/\hbar\upsilon_F\gg1)$ of width $a$. In Fig.~6 we plot the Fano factor $F$ (at~$E=V$) as function of interbarrier spacing $a$ at $V=200$~meV, $d=30$~nm, $\Delta=0$ (thin line) and $\Delta=26.5$~meV (thick line). It is clearly seen that for two gapless graphene strips kept at the Dirac point $(E=V)$ and separated by an extensive and highly doped region $(a\gg Vd^2/\hbar\upsilon_F)$ the Fano factor oscillates near the value~0.25 in accordance with analytic calculations presented in ref.~\cite{Wiener1}
. For $\Delta d/\hbar\upsilon_F>1$ and $a\gg\pi V\hbar\upsilon_F/\Delta^2$, the similar calculations yield $F = 0.5$
. For small values of $a$ the Fano factor in these cases approaches~1/3 and 1 correspondingly.

Now let us consider $n$-periodic (i.e. region $[0,\text{}(2n-1)d]$ in Fig.~\ref{fig1}(b), with $a=d$) symmetric structure. We choose $n=30$ periodic structure modeling general physical properties of a superlattice. As was already shown~\cite{Maks}, 
 depending on the potential barrier height $V$, the band structure of such SL can have more than one Dirac point located at $E=E_0$ (\ref{eq23}). In contrast to the SLs discussed, {\it e.g.}, in~\cite{Brey,Barbier}, 
the Dirac point being a prototype of the original Dirac point (situated at ${\bf k}=0$) arises at certain values of $V=V_n$:
\begin{eqnarray}\label{eq32}
V_n=\pi n\hbar\upsilon_F/d+\sqrt{(\pi n\hbar\upsilon_F/d)^2+\Delta^2}
\end{eqnarray}
 which are the solutions of equation $E_0d/\pi\hbar\upsilon_F=n$, where $n=1,\text{ }2,\dots$
\begin{figure}[t] \centering
\includegraphics*[scale=0.8]{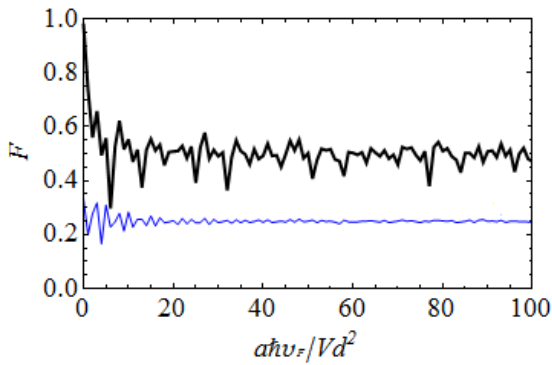}
\caption{Fano factor as a function of normalized interbarrier spacing $a$ for a double barrier system with $d=30$~nm, $E=V=200$~meV and $\Delta=0$~(thin line) and $\Delta=26.5$~meV~(thick line).}
\label{fig6}
\end{figure}
In this case the total number of Dirac points is $N_D=2n-1$. When the ratio $E_0d/\pi\hbar\upsilon_F$ is not integer, the number $N_D$ of the Dirac points symmetrically located around $k_y=0$ is given by $N_D=2[E_0d/\pi\hbar\upsilon_F]$, where $[\dots]$ denotes an integer part.

\begin{figure}[t] \centering
\includegraphics*[scale=0.8]{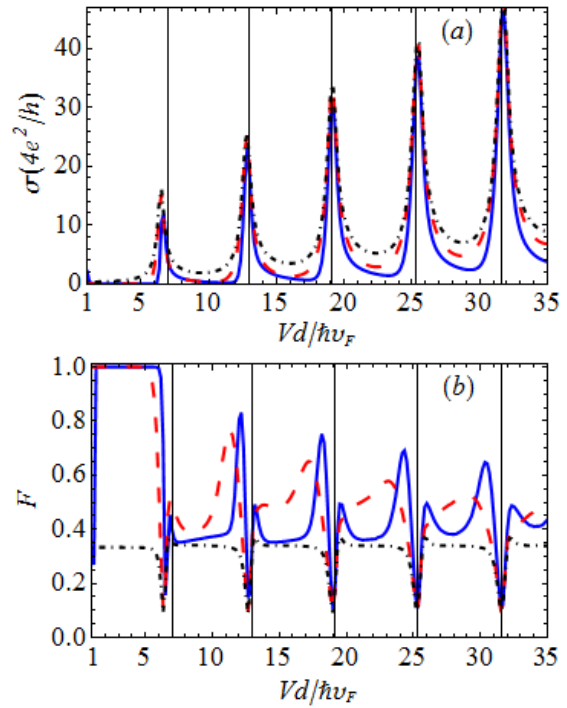}
\caption{Conductivity (a) and Fano factor (b)  at $E=E_0$ as functions of potential value $V$ for three symmetric ($a=d=30$~nm) multibarrier structures ($n=30$) characterized by different gap magnitudes in the barrier regions: $\Delta=26.5$~meV (solid line), $\Delta=10$~meV (dashed line) and $\Delta=0$ (dash-dotted line).}
\label{fig7}
\end{figure}

Fig.~\ref{fig7} shows the conductivity and the Fano factor at $E=E_0$ (\ref{eq23}) as a function of $V$ for three symmetric ($a=d=30$~nm) multibarrier structures $(n=30)$ characterized by different gap values in the barrier regions: $\Delta=26.5$~meV (solid line), $\Delta=10$~meV (dashed line), and $\Delta=0$ (dash-dotted line). Vertical lines indicate the positions of $V_n$ which weakly depend on $\Delta$ for $\Delta d/\pi\hbar\upsilon_F\ll1$ (\ref{eq32}). As seen, each a new Dirac point manifests itself as a conductivity resonance and a narrow dip in the Fano factor. Between the resonances at $\Delta=0$ the system demonstrates pseudo-diffusive behaviour $(F=1/3)$ similarly to~\cite{Fertig}. 
The existence of gapped graphene fraction in barrier regions leads to decrease of the conductivity and strongly affects the Fano factor. Independently of the gap value $\Delta$, the Fano factor $F$ of the gapped SL equals~1 almost in the whole region $\Delta<V<V_1$ that differs from $F=1/3$ for gapless graphene SL (Fig.~\ref{fig7}(b)). Such difference is caused by the qualitative distinction in the electronic spectrum of two types of the SLs in this range of $V$. For gapless SL prototype of the original Dirac point always exists in the energy spectrum. On the contrary, at $\Delta<V<V_1$ there is a minigap separating the conduction and valence minibands in the SL with $\Delta\ne0$. This results in a nearly zero transmission at $E=E_0$, and, correspondingly, $F=1$. At $V>V_1$ more complicated Fano factor dependence $F(V)$ takes place in the nonresonant domains due to nonmonotonic dependence of the transmission probability $T(E_0,V,\theta_0=0)$ in contrast to the case $\Delta=0$ when the perfect transmission occurs at $\theta_0=0$. At large heights of the potential, minimum value of $T(E_0,V,\theta_0=0)$ increases and $F(V)$ becomes smoother function of $V$ between the dips.

\section{Conclusion}

In summary, based on the transfer-matrix method, we have investigated the conductance and Fano factor as well as the angular and energy dependencies of the transmission probability for one-dimensional graphene multibarrier structures. In our study we do not consider distinction in the Fermi velocities in gapped and gapless graphene fractions that can arise, {\it e. g.}, in graphene deposited on the various substrates, or in appropriately doped graphene~\cite{Tap}.

In general case increasing the number of barriers in the considered heterostructures causes an appearance of extra peaks in transmission probability. It was found that symmetric $(a=d)$ $n$-barrier structure is perfectly transparent for some angles of incidence (\ref{eq26}) at the particle energy $E=E_0$ (\ref{eq22}). If $E_0>0$ both electronic (in the well regions) and hole (inside the barrier) states contribute to the formation of the propagating modes. In this case the positions and number of resonant peaks do not depend on the barrier number $n$. However increase of $n$ leads to the decrease of their widths, so that for $n\gg1$ the propagation of particles through 1D-graphene structure is similar to the propagation of electromagnetic waves through symmetric dielectric system composed of right- and left-handed materials.

Also, we have investigated the effect of gapped graphene fraction on the conductivity and shot noise. As expected, the inclusion of gapped graphene results in a decrease of the conductance and increase of the Fano factor. At the same time, existence of gapped-graphene regions in the structure affects the Fano factor considerably stronger than the conductivity. We have computed the conductivity and the Fano factor of the SL at $E=E_0$~(\ref{eq22}) depending on $V$. It was shown that each a new Dirac point in the SL with modulated gap manifests itself as a conductivity resonance and a narrow dip in the Fano factor similarly to a gapless SL. Between resonances the behaviour of $F(E_0,V)$ is more complicated and different from pseudo-diffusive behaviour $(F=1/3)$ typical for SL with $\Delta=0$. It was also shown, that irrespective of the gap value $\Delta$ in the range of potential values $\Delta<V<V_1$ the Fano factor $F(E_0,V)$ equal~1 for gapped SL unlike value 1/3 for a gapless SL.

\appendix
\section{}

Since the multibarrier structure consists of two kinds of graphene strips, there are two different matrices $F_d$ and $F_a$
\begin{eqnarray}\label{eqa1}
F_d=\frac{1}{\cos\theta}\pmatrix{\cos(\beta-\theta) & \frac{i\sigma}{\delta}\sin\beta \cr
i\sigma\delta\sin\beta & \cos(\beta+\theta)},
\end{eqnarray}
\begin{eqnarray}\label{eqa2}
F_a=\frac{1}{\cos\theta_0}\pmatrix{\cos(\beta_0-\theta_0) & i\sigma_0\sin\beta_0 \cr
i\sigma_0\sin\beta_0 & \cos(\beta_0+\theta_0)}
\end{eqnarray}
with $\beta=kd\cos\theta$, $\beta_0=k_0a\cos\theta_0$, $\sigma_0=sgnE$, $\sigma=sgn(E-V+\Delta)$, $k_0=|E|/(\hbar \upsilon_F)$, $k=\frac{\sqrt{(V-E)^2-\Delta^2}}{\hbar \upsilon_F}$.

As was noted above, we suppose that $k^2>k_y^2$ where $k_y=k_0\sin\theta_0=k\sin\theta$. Let the superlattice contain $n$ barrier regions of width $d$. Then it is easy to see from Eq.~(\ref{eq10}), that $M^{(n)}=GS^{n-1}L_0$, where $G=L_0^{-1}F_d$ and $S=F_aF_d$. Using Eqs.~(\ref{eq5}), (\ref{eqa1}) and (\ref{eqa2}) after some algebra we obtain
\begin{eqnarray}\label{eqa3}
&&G=\frac{1}{\sqrt 2\cos\theta_0\cos\theta}\pmatrix{g_{11} & g_{12} \cr
g_{21} & g_{22}}, \nonumber\\
&&S=\pmatrix{s_{11} & is_{12}\cr is_{21} & s_{22}},
\end{eqnarray}
\begin{eqnarray}\label{eqa4}
&&g_{11}=\cos(\beta-\theta)\exp(-i\theta_0)+i\sigma\sigma_0\delta\sin\beta,\nonumber\\
&&g_{12}=\sigma_0\cos(\beta+\theta)+i\frac{\sigma}{\delta}\sin\beta \exp(-i\theta_0),\nonumber\\
&&g_{21}=\cos(\beta-\theta)\exp(i\theta_0)-i\sigma\sigma_0\delta\sin\beta,\nonumber\\
&&g_{22}=i\frac{\sigma}{\delta}\sin\beta \exp(i\theta_0)-\sigma_0\cos(\beta+\theta),
\end{eqnarray}
and
\begin{eqnarray}\label{eqa5}
&&s_{11}=\frac{\cos(\beta_0-\theta_0)\cos(\beta-\theta)-\sigma\sigma_0\delta\sin\beta\sin\beta_0}{\cos\theta\cos\theta_0},\nonumber\\
&&s_{12}=\frac{\frac{\sigma}{\delta}\sin\beta\cos(\beta_0-\theta_0)+\sigma_0\sin\beta_0\cos(\beta+\theta)}{\cos\theta\cos\theta_0},\nonumber\\
&&s_{21}=\frac{\sigma_0\sin\beta_0\cos(\beta-\theta)+\sigma\delta\cos(\beta_0+\theta_0)\sin\beta}{\cos\theta\cos\theta_0},\nonumber\\
&&s_{22}=\frac{-\frac{\sigma\sigma_0}{\delta}\sin\beta_0\sin\beta+\cos(\beta_0+\theta_0)\cos(\beta+\theta)}{\cos\theta\cos\theta_0}.\nonumber\\
&&
\end{eqnarray}
The above expressions allow us to find the matrix element $M_{22}^{(n)}$ determining the transmission (\ref{eq19}). Thus for a single barrier $(n=1)$ one obtains
\begin{eqnarray}\label{eqa6}
&&M_{22}^{(1)}=\cos\beta+i\sin\beta(\tan\theta\tan\theta_0-\nonumber\\
&&-\sigma\sigma_0(\delta+1/\delta)/2\cos\theta\cos\theta_0).
\end{eqnarray}
The value $n=2$ corresponds to the transmission of electron through a symmetrical double barrier structure. In this case
\begin{eqnarray}\label{eqa7}
&&Re (M_{22})=\frac{1}{2\cos^2\theta\cos\theta_0}\{\cos^2(\beta-\theta)\cos(\beta_0-\theta_0)+\nonumber\\
&&+\cos^2(\beta+\theta)\cos(\beta_0+\theta_0)-\sigma\sigma_0(\delta+1/\delta)\sin 2\beta\sin\beta_0\times\nonumber\\
&&\times\cos\theta-2\sin^2\beta\cos\beta_0\cos\theta_0\},
\end{eqnarray}
\begin{eqnarray}\label{eqa8}
&&Im (M_{22})=\frac{1}{2\cos^2\theta\cos^2\theta_0}\{\sin\theta_0[\cos^2(\beta-\theta)\times\nonumber\\
&&\times\cos(\beta_0-\theta_0)-\cos^2(\beta+\theta)\cos(\beta_0+\theta_0)+\nonumber\\
&&+2\sin^2\beta\sin\beta_0\sin\theta_0-2\sigma\sigma_0(\delta+1/\delta)\sin^ 2\beta\sin\beta_0\sin\theta]-\nonumber\\
&&-2\sin\beta_0\cos(\beta+\theta)\cos(\beta-\theta)-\sigma\sigma_0(\delta+1/\delta)\sin\beta\times\nonumber\\
&&\times[\cos(\beta-\theta)\cos(\beta_0-\theta_0)+\cos(\beta+\theta)\cos(\beta_0+\theta_0)]+\nonumber\\
&&+(\delta^2+1/\delta^2)\sin^2\beta\sin\beta_0\}.
\end{eqnarray}

\section*{Acknowledgments}

We are grateful to Dr.~Burdov for his interest in this investigation and for helpful remarks. This work was supported by the Russian Foundation for Basic Research (Grant No 13-02-00784)

\end{document}